# Two-spin-multiplexed optoacoustic light storage in chiral photonic crystal fiber

Xinglin Zeng[1,4*#], Linqiao Gan[1,2#], Jesús Humberto Marines Cabello[1,2], Olivia Saffer[1,2] and Birgit Stiller[1,3*]

[1]*Max Planck Institute for the Science of Light, Staudtstr. 2, 91058 Erlangen, Germany*
[2]*Department of Physics, Friedrich-Alexander Universität, Staudtstr. 7, 91058 Erlangen, Germany*
[3]*Institute of Photonics, Leibniz University Hannover, Welfengarten 1A, 30167 Hannover, Germany*
[4]*Russell Centre for Advanced Lightwave Science, Shanghai Institute of Optics and Fine Mechanics and Hangzhou Institute of Optics and Fine Mechanics, Chinese Academy of Sciences, Shanghai 201800, China*

*\*E-mail: xinglin.zeng@r-cals.com, birgit.stiller@mpl.mpg.de*
[#]*Xinglin Zeng and Linqiao Gan contributed equally to this work.*

**Abstract:** The ability to coherently store and manipulate optical information across multiple degrees of freedom is a central requirement for scalable quantum information processing and multidimensional quantum computing. While polarization- and space-division multiplexing have substantially increased the capacity of classical optical systems, their extension to coherent and reconfigurable photonic memories remains a key challenge. Here we demonstrate a two-spin-channel-multiplexed photonic memory based on chiral stimulated Brillouin scattering in a chiral photonic crystal fiber. Exploiting the intrinsic preservation of circular polarization in the chiral photonic crystal fiber, left- and right-circularly polarized modes serve as two orthogonal and independent storage channels. Multiple optical data pulses can be selectively or simultaneously stored and retrieved by simply controlling the polarization states of the write-read pulses. The storage time is continuously tunable, and the underlying Brillouin process preserves coherence and channel orthogonality. The result establishes chiral Brillouin scattering as an effective mechanism for spin-channel-multiplexed optoacoustic light storage, providing a robust and scalable platform for multidimensional photonic memories. It also open new opportunities for classic and quantum information processing, reconfigurable quantum networks, and hybrid light–matter interfaces based on coherent acoustic excitations.

# 1. Introduction

In optical fiber systems, the demand for increasing information capacity is driven not only by the rapid growth of classical data traffic but also by emerging quantum information technologies[1–4]. Conventional multiplexing approaches such as wavelength-division multiplexing (WDM), while highly effective, are approaching their fundamental capacity limits imposed by nonlinear effects[5,6]. This has stimulated the development of advanced multiplexing strategies, including polarization-division and space-division multiplexing, which exploit additional optical degrees of freedom to enhance information density[7–11]. Beyond classical communications, these techniques provide a natural platform for multidimensional quantum information processing, where polarization and spatial modes enable high-dimensional quantum state encoding[12–15]. Scalable quantum computing and quantum networks further require coherent and reconfigurable photonic memories capable of preserving multiple degrees of freedom simultaneously, motivating the exploration of multiplexed photonic storage architectures for next-generation quantum technologies[16–18].

Stimulated Brillouin scattering (SBS) is one of the most widely utilized physical effects in photonic memory systems[19–21]. It enables the coherent transfer of optical information onto acoustic waves, which propagate at speeds approximately five orders of magnitude slower than light. Such significant velocity difference provides a natural and efficient means for the temporary storage and controlled retrieval of optical signals. Beyond that, SBS has recently emerged as a key physical platform for implementing optoacoustic recurrent operators in high-performance optical neuromorphic computing systems[22]. To date, Brillouin light storage has demonstrated the ability to store data encoded in amplitude[19], frequency[23], and phase[24]. However, the polarization and spatial degrees of freedom, which are essential for multi-dimensional photonics and classic/quantum information processing, have yet to be systematically explored in Brillouin-based memory architectures.

Here, we demonstrate a two-spin-multiplexed (TSM) Brillouin light storage in chiral photonic crystal fiber (PCF) that features a circular birefringence and vortex birefringence, as well as a large-air-filled silica micro-structure that enables strong optoacoustic overlap. Building upon the conventional Brillouin light storage, we implement a more advanced scheme, in which two data pulses—each encoded with a distinct and mutually orthogonal spin state (i.e., circular polarization)—are injected into the fiber. Correspondingly, two pairs of control pulses (each pair has one writing pulse and one reading pulse) are launched to store and retrieve the data. As dictated by the cross-spin SBS effect in chiral PCF[25], the data pulses can be selectively or simultaneously written and readout by appropriately tuning the spin state of each control pulse pair. By leveraging the inherent chirality of the PCF, our approach not only enhances the selectivity and capacity of the optoacoustic light storage but also provides a new degree of freedom for manipulating light-matter interactions. The results highlight the promising role of chiral PCF and spin-state-multiplexing in advancing photonic memory technologies, with potential applications in optical buffering, quantum information processing, and high-speed all-optical signal routing.

## 2. Results

**Brillouin scattering in large air-filled chiral PCF**

The chiral PCF, which is fabricated by rotating the preform during the fiber drawing, have the ability of robustly transmitting spin angular momentum (SAM) and orbital angular momentum (OAM)[26]. Its advent has enabled many new and interesting studies on the optical nonlinear effects (including Brillouin scattering) in the presence of chirality and related applications. In chiral PCF it is found, both experimentally and by numerical modelling, that the fields carry pure spin states $s$ ($s = +1$ for left-circular polarization (LCP) and $-1$ for right-circular polarization (RCP)), under which circumstances the topological charge $\ell^{(m)}$ of OAM (the number of on-axis discontinuities for fields evaluated in Cartesian coordinates) is linked to the azimuthal order by $\ell^{(m)} = \ell_A^{(m)} - s$, where m is the harmonic order of helical Bloch mode in chiral PCF. Here, we only consider the modes in the first Brillouin zone and therefore use the shorthand $[\ell, s]$ to denote the parameters of an eigenmode in chiral PCF, where for ease of notation the principal topological order is defined as $\ell^{(0)} \equiv \ell$. Fig. 1(a) shows a 3D image of a Y-shaped-core chiral PCF. Fig. 1(b) shows the scanning electron micrograph (SEM) of the cross-section of the fabricated chiral PCF. The measured geometric parameters include satellite core size of 1.56 μm, hollow channel diameter of 1.9 μm, spacing between adjacent hollows of 1.78 μm, and a twist period of 5 mm. The design of three off-axis satellite cores allows the fiber to have a circular birefringence of $8\times10^{-6}$ and topology birefringence of $1.5\times10^{-3}$, which make the fiber stably preserving spin states and vortex states over long distance. The fiber loss at 1550 nm, measured by cut-back method, is ~0.195 dB/m for $\ell = 0$ modes and 0.256 dB/m for $\ell = 1$ modes. Fig. 1(c) shows the measured near field distribution of $[0,+1]_1$ (i.e., first-order radial mode with LCP), $[+1,+1]$ (i.e., left-handness OAM with LCP) and $[0,+1]_2$ (i.e., second-order radial mode with LCP), at the output of 1.5 m-length fiber. We note that the subscript 1 or 2 represent the radial order of circularly polarized mode. The left panel in Fig. 1(d) shows the calculated strain field densities of acoustic mode that mediate the Brillouin scattering of all three optical modes at 1550 nm. Owing to the phase-matching condition, the same acoustic mode of different frequency is involved in each Brillouin scattering process. The right panel in Fig. 1(d) shows another acoustic mode that works for $[0,\pm1]_1$ optical mode. Fig. 1(e) shows the measurements of polarization maintaining ability after the $[0,\pm1]_1$ and $[+1,\pm1]$ modes propagate along the 1.5 m length of the fiber, and the modulus of the Stokes parameter $|S_3|$ are higher than 0.93 at the output, showing very good preservation of SAM and OAM. Fig. 1(f) shows the measured circular birefringence ($B_c$) between $[0,+1]_1$ and $[0,-1]_1$ modes and between $[+1,+1]_1$ and $[+1,-1]_1$, as a function of wavelength. The optical activity of the fiber enables the linear polarization states to rotate while changing the laser wavelength. The rotation angle (in rad/m) is related to the $B_c$ by $2\pi B_c/\lambda$, where λ is the wavelength.

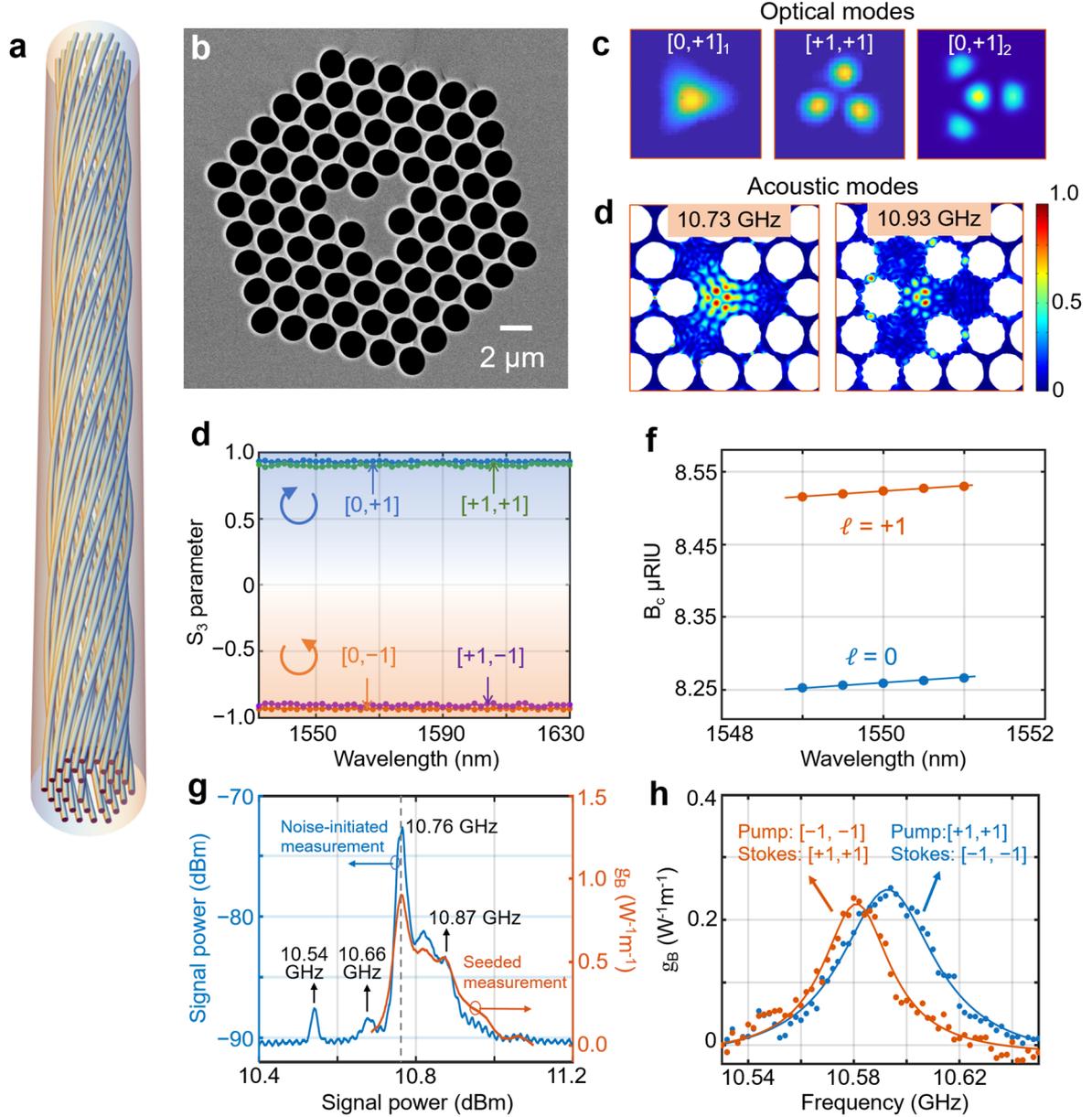

**Fig. 1.** (a) 3D illustrative image of Y-shape chiral PCF. (b) Scanning electron micrograph (SEM) of chiral PCF. (c) Measured near-field distribution of Gaussian-like $[0,+1]_1$ mode, $[+1,+1]$ mode and $LP_{02}$-like $[0,+1]_2$ mode at the output of 1.5 m-long fiber. (d) The simulated strain field density of two acoustic modes (10.733 GHz and 10.932 GHz) having large overlap integral with fundamental optical modes. (e) Measured Stokes parameters $S_3$ at the output of 1.5 m-long fiber when $[0,+1]_1$, $[0,-1]_1$, $[+1,+1]$ and $[-1,-1]$ modes are launched. All values are higher than 0.93 at the output, showing very good preservation of SAM and OAM. (f) Measured circular birefringence $B_c$ between two $\ell = 0$ modes (blue dots) and between $\ell = +1$ modes (orange dots), and their theoretical fittings (solid lines), as functions of wavelength. (g) Blue: Noise-initiated spontaneous Brillouin spectrum generated by pumping $[0,+1]_1$ at 1550 nm into the chiral PCF. Orange: Brillouin gain spectrum measured by pump-seed setup. The pump and seed have opposite circular polarization states. The highest Brillouin peaks is located at 10.76 GHz, which is in good agreement with theoretical value of 10.733 GHz. Many hybrid acoustic modes exists around, making the gain spectrum much broader. The peaks at 10.54 GHz and 10.66 GHz originate from the Brillouin scattering of $[0,+1]_2$ and $[+1,+1]$ modes, which come from the mode crosstalk of $[0,+1]_1$ mode. (h) Measured Brillouin gain factor as functions of pump-Stokes frequency difference, when the pump is $[+1,+1]$ and $[-1,-1]$.

The blue curve in Fig. 1(g) shows the noise-initiated spontaneous Brillouin spectrum generated by pumping [0,+1]$_1$ mode into the fiber. The highest Brillouin peak is located at 10.76 GHz, which is in good agreement with theoretically values of 10.73 GHz. The high air-filling fraction and small core size of the fiber lead to the coupling between the longitudinal and shear components of the acoustic waves, which enhances the excitation of multiple acoustic modes during Brillouin scattering and, as a result, the Brillouin spectrum exhibits parasitic peaks and a broadened bandwidth. The strain field density of acoustic modes of highest Brillouin peak at 10.73 GHz and a side-peak at 10.93 GHz are shown Fig. 2(d). For details on the spontaneous Brillouin measurement and Brillouin gain coefficient measurement please see the Method Sec. 1 and 2. The orange curve in Fig. 1(g) shows the Brillouin gain spectrum measured by launching [0,+1]$_1$ pump and [0,−1]$_1$ seed in opposite direction. In chiral PCF, the angular momentum conservation dictates that the pump and Stokes have opposite topological charges and spin orders. An additional peak appears at 10.54 GHz, and the measurement using near-field scanning Brillouin analyzer (NBA) indicates that it comes from the Brillouin scattering of circular polarization (CP) [0,+1]$_2$ mode, which in generated through intermodal linear coupling from [0,+1]$_1$ mode. For details on NBA results please see Supplementary Materials S1. A smaller peak at 10.67 GHz is also observed, arising from the Brillouin scattering of CP vortex modes that are produced via intermodal coupling from the [0,+1]$_1$ mode, as confirmed again by the NBA detection system. Fig. 1(h) shows the measured Brillouin gain factor as a function of pump-Stokes frequency difference when the pump is [+1,+1] (blue curve) and [−1,−1] (orange curve), at the wavelength of 1550 nm. The Brillouin peak frequencies in both cases are 10.65 GHz and 10.664 GHz, indicating a 14 MHz shift. Such frequency differences are in good agreement with theoretical values of 11.5 MHz, which are calculated from $2\Delta n_{\text{eff}} V_{ac}/\lambda$. Here, the $\Delta n_{\text{eff}}$=0.0015 is the index difference between [+1,+1] and [−1,−1]. The $\lambda$ is the optical wavelength in vacuum, and $V_{ac}$ is the velocity of longitudinal acoustic waves, which is approximately 5970 m/s for backward Brillouin scattering.

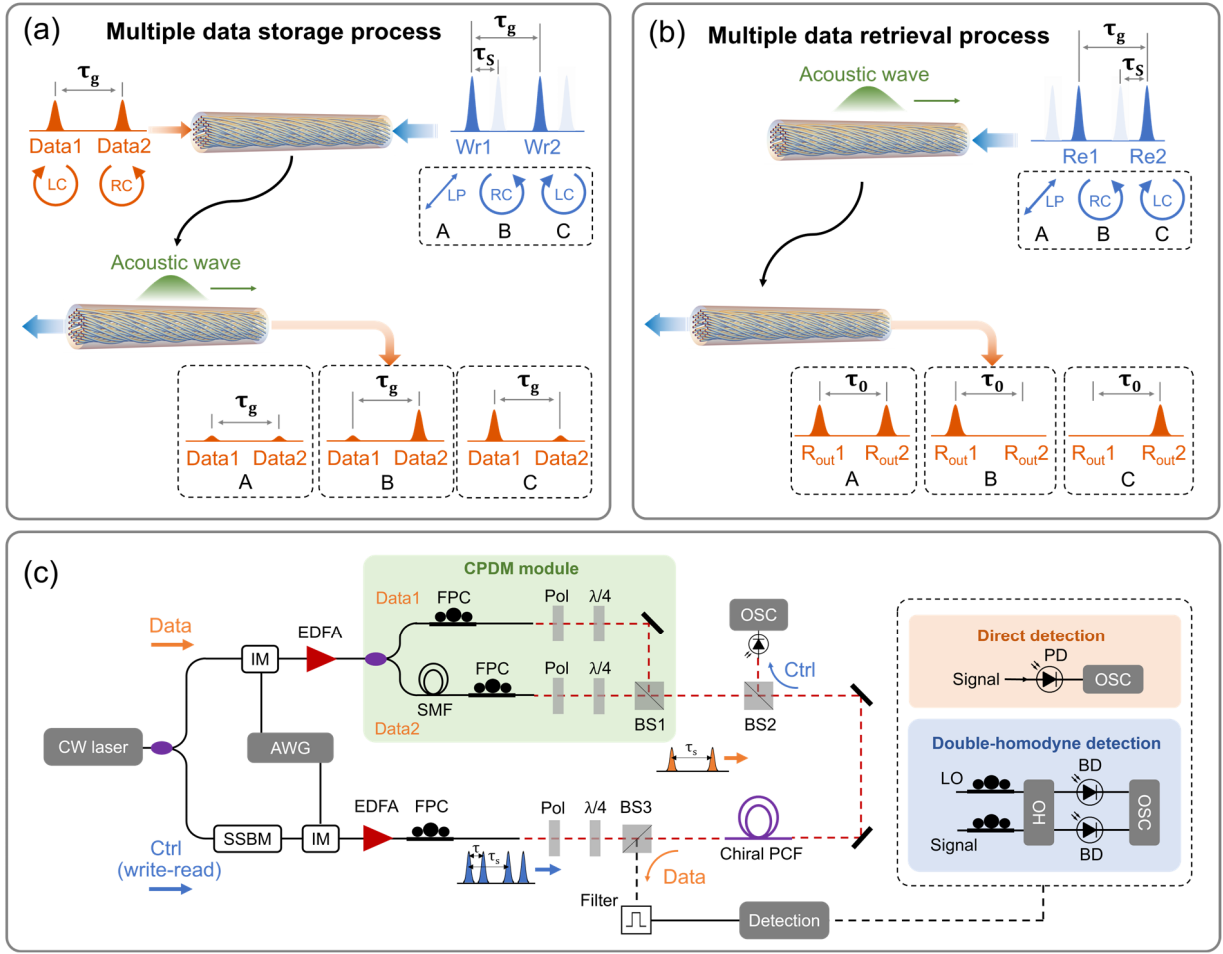

**Fig. 2** (a) Two-spin-multiplexed (TSM) data-storing process: the information encoded in data pulses is stored into acoustic waves through interaction with two strong counter-propagating write pulses. Simultaneous or selective storage of the data pulses is achieved by tuning the polarization states of the write pulses to LP or CP, manifested as depletion of the data pulses. (b) TSM data retrieval process: After a controlled storage time, two read pulses following their respective write pulses deplete the acoustic phonons, converting the data information back to optical domain. Like the writing process, whether the data is retrieved simultaneously or selectively depends on the polarization state of the read pulses. (c) Experimental setup of TSM light storage with chiral PCF. CW, continuous wave; Ctrl, control; SSBM, single sideband modulator; IM, intensity modulator; AWG, arbitrary wavefunction generator; EDFA, erbium-doped fiber amplifier; FPC, fiber polarization controller; SMF, single mode fiber; Pol, polarizer; λ/4, quarter-wave plate; BS, beam splitter; PD, photodiode; BD, balanced detector; OSC, oscilloscope.

**Two-spin-multiplexed (TSM) optoacoustic light storage**

Next, we demonstrate two-spin-channel-multiplexed (TSM) optoacoustic light storage, by exploiting the SBS effect in a chiral photonic crystal fiber (PCF). As a proof-of-concept, the $[0,+1]_1$ and $[0,-1]_1$ modes—fundamental modes with two spin states—are employed as two independent storage channels. For simplicity, these two modes are hereafter referred to as LCP and RCP modes. Fig. 2(a) and 2(b) schematically illustrate the operating principle of the TSM Brillouin light storage system, showing both the data-storage and retrieval (read-out) processes. In the scheme, two orthogonal LCP and RCP states are independently encoded onto two optical data pulses that are temporally separated by an interval $\tau_g$. Each data pulse counter-propagates against a pair of strong optical write-read pulses, which are responsible for the storage and retrieval of data, respectively. The storage time $\tau_s$ can be precisely and continuously controlled by adjusting the time interval between the write and read pulses, offering flexible and dynamic reconfigurable memory operation.

As shown in Fig. 2(a), during the data-storage stage, two strong optical write pulse (Wr1 and Wr2) at a lower optical frequency $f_w$ are launched into the fiber in the direction opposite to that of two data pulses, which have higher frequency $f_d$. Two data pulses are in two orthogonal spin states, representing independent signal channels, while write pulses are in one spin state. As the write and data pulses encounter each other, the SBS effect between them leads to the coherent generation of acoustic phonons in the fiber, and as a result, the data pulses will be strongly depleted via the Brillouin Stokes process. The frequency of the generated acoustic wave is the Brillouin frequency shift $f_B = f_w - f_d$. There are two types of writing process: simultaneous and selective writings. When the two write pulses are in a linear polarization (LP) that is a coherent combination of two spin states, both will couple to the respective data pulses and deplete them simultaneously (case A in Fig. 2(a)). However, when the write pulses are in one spin state (LCP or RCP), only the data pulse with the opposite spin state is selectively depleted and stored, while the one with the same spin as the write pulse remains unaffected due to the spin-selective nature of the Brillouin interaction (cases B and C in Fig. 2(a)). Once the information from the optical data pulses is transferred to the acoustic wave, it can be retrieved using read pulses, as shown in Fig. 2(b). Two strong read pulses (Re1 and Re2), delayed by a time interval $\tau_s$ (i.e., the storage time) with respect to their corresponding write pulses, are launched into the fiber. These read pulses share the same polarization states as the write pulses and interact with the acoustic waves through the Brillouin anti-Stokes process. Like the writing stage, the data retrieval is governed by the polarization states of the read pulses. When the read pulses are linearly polarized (LP), both Data1 and Data2 are retrieved, as illustrated in case A in Fig. 2(b). In contrast, when they are in CP state, only the data encoded in the opposite CP state is retrieved, while the other that in the same CP state cannot be retrieved, as illustrated in cases B and C of Fig. 2(b).

Figure 2(c) shows the experimental setup for the proof-of-concept demonstration. A continuous wave (CW) laser is divided into two paths: one is for the generation of data pulses and the other is for the generation of write-read pulses. The latter one is frequency down-shifted by a Brillouin frequency $f_B$ by using a single sideband modulator (SSBM). The concept of TSM (green region) is realized by splitting the data pulses into two paths and then recombining them using a beam splitter (BS1), with a time delay between the two data channels is introduced in one path using a single-mode fiber. The spin state of each data path is separately tuned by two pairs of polarizer and quarter-wave plate. For details on the experimental setup please see Method Sec. 4.

We first demonstrate light storage of a single data pulse with spin state, using a simplified experimental setup (see Method Sec. 3). The storage medium is a 1.5 m-long chiral PCF. The input data pulse has a duration of 1.8 ns and a peak power of 10 W. The write and read pulses also have durations of 1.8 ns, with peak powers of 56 W. Fig. 3(a) shows the measured data and readout pulses. The cross-spin backward SBS effect in chiral PCF leads to significant data depletion of exceeding 40% and enables data retrieval with an efficiency greater than 9%. The inset of Fig. 3(a) shows the write and read pulses separated by a time interval of $\tau_s = 3$ ns, resulting in the same delay between the data and retrieved pulses. In contrast, when the data and control pulses have opposite spin, no storage or retrieval occurs, as shown in Fig. 3(b). The storage time can be continuously tuned by adjusting the delay between the write and read pulses, as shown in Fig. 3(c), with the maximum storage duration limited by the acoustic lifetime. From the exponential decrease of the retrieval efficiency, an acoustic decay time of 4 ns is estimated by an exponential fitting and is confirmed by the measurement of the Brillouin gain linewidth, which is ~40 MHz (Fig. 1(g)).

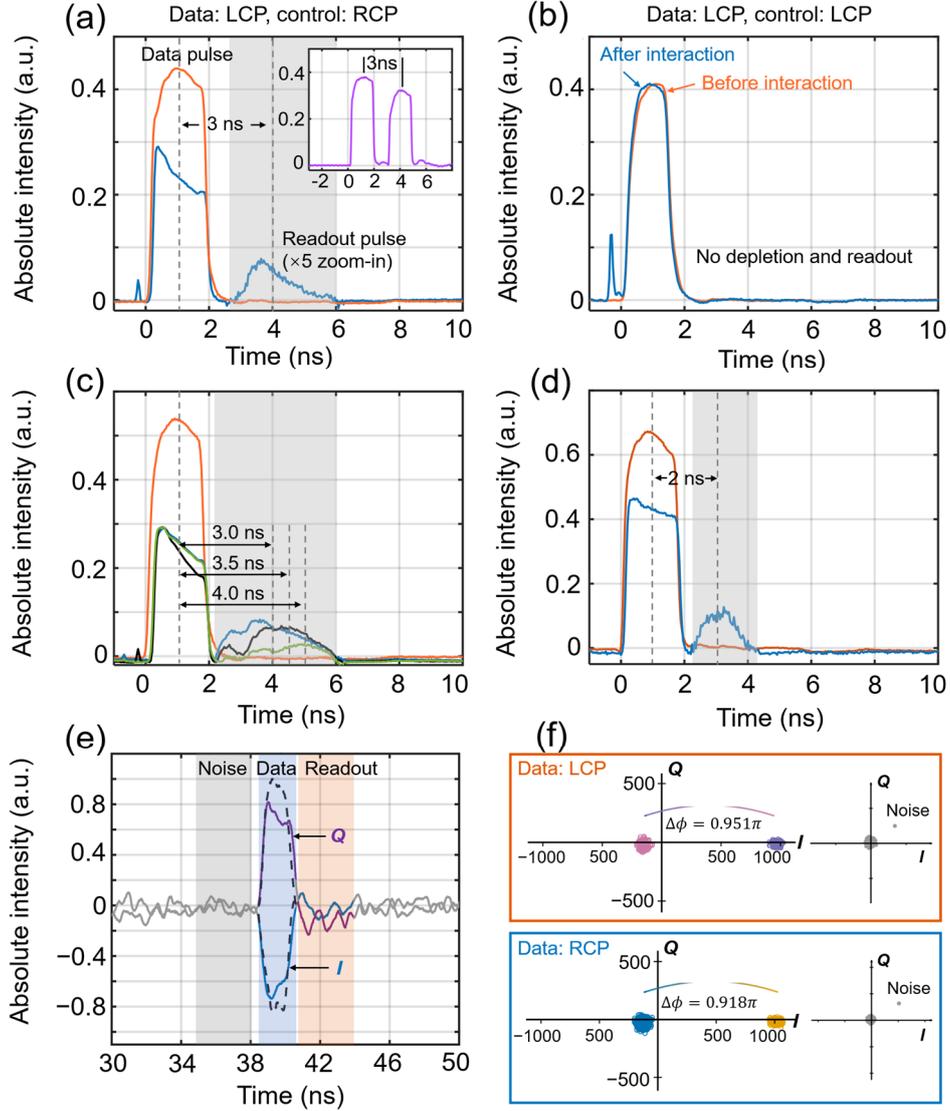

**Fig. 3** (a) Measured initial data pulses (orange) and readout pulses (blue) when the data is in LCP and write-read is in RCP. The inset shows the write-read pulses. The frequency between data and write-read is 10.76 GHz. (b) Same measurement, but both data and write-read are in LCP. No depletion and readout are observed in this case. (c) Retrieved data pulses after different storage time. (d) Data depletion and retrieval when the frequency difference between data and write-read is set as 10.87 GHz. (e) Time-resolved in-phase $I$ (blue) and quadrature $Q$ (purple) components of depleted data pulses and readout pulses, measured via double balanced homodyne detection (BHD). (f) Upper one: Constellation plot for readout process and noise, when the data is LCP and write-read is RCP. Lower one: same as upper one, except that the data is RCP and write-read is LCP. The measured differential phase are $\Delta\varphi \approx 0.951\pi$ and $0.918\pi$ in both cases, showing stable phase correlation of $\pi$ in optoacoustic storage process.

It has been shown in Fig. 1(g) that when injecting a fundamental mode into the fiber, more than one acoustic mode is excited, leading to a wideband Brillouin spectrum. Therefore, besides the main Brillouin peak at 10.76 GHz, the other peaks are worthy to be examined, to check their potential suitability for light storage. Fig. 3(d) shows the results of write and read at frequency of 10.87 GHz, which is the most right-side Brillouin peak in the gain spectrum. The data pulse is in LCP and write-read pulses are in RCP. The data depletion and readout generation are clearly observed in the figure, with a depletion rate exceeding 35% and a readout efficiency greater than 7.6%. As before, when the data and write-read pulses share the same circular polarization state, no depletion or signal retrieval is observed.

Next, we measured the phase differences between original data and readout pulses under different polarization states (LCP and RCP) by using double balanced homodyne detection (BHD) scheme, which consists of a 90° optical hybrid, two 2.5-GHz balanced photodetectors, and a high-speed oscilloscope. The BHD setup is shown in the right corner of Fig. 2(c). Fig. 4(e) shows the in-phase $I(t)$ and quadrature $Q(t)$ traces measured by double homodyne system simultaneously. For phase analysis, the areas under the measured $I(t)$ and $Q(t)$ traces (AuC$_I$, AuC$_Q$) of the depleted data and readout data pulses are calculated over their respective pulse windows of 2.15 ns and 3.2 ns. The initial data has {00} phase. The readout data exhibit an intrinsic phase shift of π (~0.951π for LCP data and RCP write-read configuration, ~0.918π for RCP data and LCP write-read configuration) relative to the initial data[27], corresponding to a symbol transition from {00} to {10} in the QPSK constellation, as shown in Fig. 3(f). For comparison, an identical 3.2 ns time window is selected from a noise-only region, where the measured phases deviate significantly from $π$ and exhibit random fluctuations. Therefore, the cross-spin light storage in chiral PCF maintains the intrinsic coherence, preserving the phase relation between the data pulses and its retrieved signals, which suggests the potential for polarization-resolved phase encoding in Brillouin-based optical memory, and enables potential applications in quantum photonics. For details on the phase measurement with BHD, please see Supplementary Materials S2.

We next utilize the setup in Fig. 2(c) to demonstrate the write-read process on TSM data pulses. The time interval $\tau_g$ between the two data pulses is set to 25 ns and 97 ns, corresponding to propagation delays in 5 m-long and 20 m-long single-mode fibers (SMFs), respectively. These intervals also define the delays between the two pairs of write-read control pulses, as shown in Fig. 4(a) and 4(b). The storage time $\tau_s$ (time delay between write and read pulses) is fixed at 3 ns for both pulse pairs. Fig. 4(c) shows the results of data storage and retrieval when Data1 is in LCP and Data2 is in RCP. As expected, when the write-read pulses are in RCP, as shown in the upper panel, only the Data1 pulse is selectively depleted and retrieved, with a data depletion rate of 64% and a read-out efficiency of 12.2%, while the Data2 pulse keep unchanged; on the contrary, when the control is in LCP, the Data2 pulse is selectively stored and retrieved, with a depletion rate of 58.1% and a read-out efficiency of 10.8%, and the Data2 remains unaffected, as shown in the central panel; at last, when the control pulse is in LP state that consists of LCP and RCP, both data pulses are stored and retrieved, as shown in the bottom panel, with depletion and read-out efficiencies of 59.2% and 9.6% for Data1, and 73.4% and 9.8% for Data2, respectively. Fig. (4d) shows the same experiment, but the time interval between two data pulses is 97 ns. It is important to note that, although the two data pulses are temporally separated, they can still interfere with each other, as they come from one laser source and the pulse modulation is not perfect. Their residual light in the "dark regions" of pulse trains interferes during the Mach-Zehnder pulse combination process (the green area in Fig. 2), making the intensity of both data pulses drift over time. To mitigate this effect, each trace was averaged 1000 times in the measurement. Nevertheless, the depletion and read-out efficiencies of two data pulses differ in efficiency from the results of single data storage experiment. Although an ideal configuration would employ two independent laser sources to generate the two data pulses separately, a single laser source is used here to simplify the experimental setup for this proof-of-concept demonstration of TSM light storage.

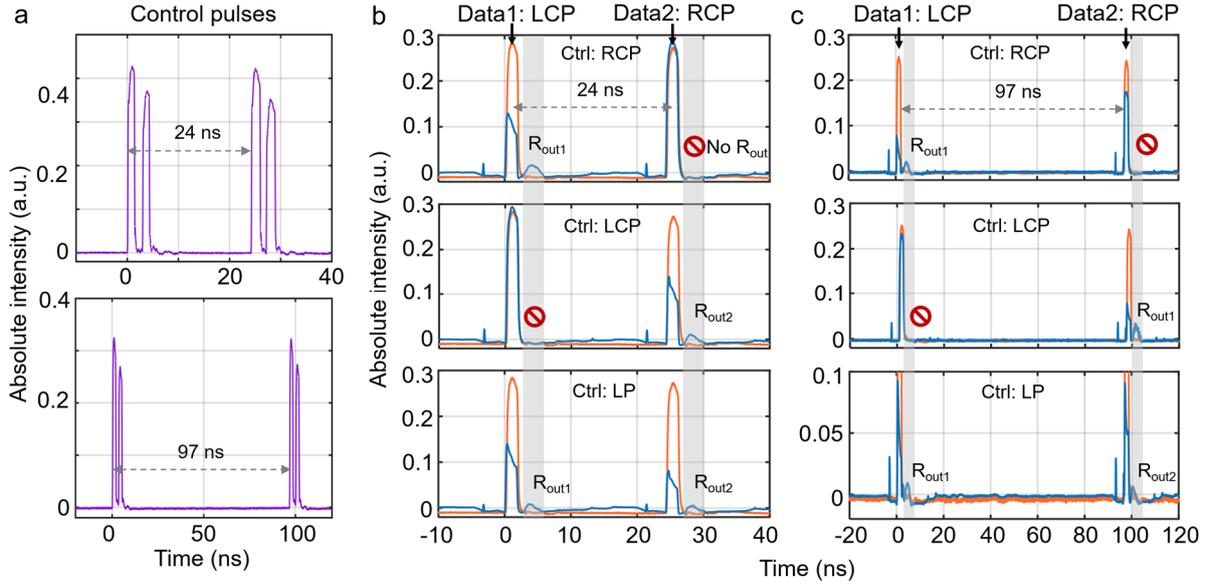

**Fig. 4** Recorded write-read pulses and data pulses in the two-spin-channel-multiplexed light storage experiment. (a) Two pairs of write-read pulses separated by 24 ns (upper one) and 97 ns (lower one). Each pair works for the storage and retrieval of one data. (b) The recorded pulses showing selectively (upper two) and simultaneously (bottom one) write-read two multiplexed data of 24 ns time gap. (c) Same as (b), but the time gap between two data pulses is 97 ns.

## 3. Discussion and Conclusions

Chiral PCF shows unique properties of chirality and is capable of robustly preserving SAM and OAM, which allows the scaling of optoacoustic memory to a multi-dimensional fashion. In the paper, we have demonstrated a two-spin-channel-multiplexed optoacoustic light storage scheme based on the cross-spin SBS effect in chiral PCF. $s=1$ and $s=-1$ (i.e., LCP and RCP) optical modes are employed as two orthogonal and independent memory channels within a single fiber platform. This enables selective as well as simultaneous storage and retrieval of optical data streams by purely polarization-based control. The demonstrated storage process exhibits continuously tunable delay, coherence preservation, and strict spin selectivity governed by angular-momentum conservation. Using balanced homodyne detection, we further confirm that the phase relation between the stored and retrieved signals is maintained, highlighting the suitability of this platform for phase-sensitive operations. These features are essential for multidimensional photonic memories and represent a significant extension of Brillouin-based light storage beyond conventional amplitude-, frequency-, and phase-only encoding schemes, which is crucially for classic and quantum information processing.

The demonstrated two-spin-channel-multiplexed Brillouin light storage highlights the unique role of chirality in mediating spin-selective optoacoustic interactions. In contrast to conventional silica fibers, where polarization and spatial degrees of freedom are generally fragile and strongly affected by perturbations[28], the chiral photonic crystal fiber employed here provides intrinsic protection of circular polarization states and angular momentum. This robustness is essential for enabling reliable channel orthogonality and repeatable light–sound conversion, and it distinguishes the present platform from previously reported Brillouin memory schemes based on standard fibers[29] or planar waveguides[30].

While the current demonstration focuses on two spin channels in a proof-of-concept configuration, the underlying principle is inherently scalable. It can be further extended toward higher storage capacity by incorporating both SAM and OAM channels. Realizing such a scheme requires sufficiently large Brillouin gain for OAM-carrying optical modes. Although in the present work the measured Brillouin gain coefficient for OAM modes is relatively modest (~0.25 $W^{-1}$ $m^{-1}$), it is worth noting that Brillouin gain can be enhanced by up to two orders of magnitude when soft-glass platforms—such as chalcogenide glasses $As_2Se_3$ or $As_2S_3$—are employed instead of silica[31]. While the fabrication of chiral photonic crystal fibers based on these materials remains technically challenging, their use would enable highly efficient spin- and orbital-channel-multiplexed light storage within much shorter fiber lengths. Future implementations could combine spin/orbital-channel-multiplexing with wavelength-, time-, or phase-multiplexing to realize high-capacity, multi-dimensional optoacoustic memories. Together with advances in low-loss chiral fiber fabrication and integrated polarization control, this work lays the foundation for a new class of optoacoustic devices that exploit chirality as a functional degree of freedom.

In addition, the optoacoustic interaction strength can be further increased by operating with data and write–read pulse trains at lower repetition rates, thereby allowing the local Brillouin response to fully build up. This approach would enhance the effective Brillouin coupling even in silica-based chiral PCFs, rendering orbital-channel-multiplexing more feasible within the current platform. Together, these considerations indicate a clear pathway toward scalable, high-dimensional optoacoustic memories that exploit the combined degrees of freedom of spin and orbital angular momentum.

Finally, besides the classic and quantum information applications, the present system is well suited for neuromorphic and analog photonic computing[32]. Brillouin interactions have recently been identified as optoacoustic building blocks for recurrent operators[22,33] and photonic neural network[34], owing to their intrinsic nonlinearity, delay, and memory effects. The introduction of spin-multiplexed channels adds an additional layer of parallelism, enabling multi-dimensional state spaces and potentially richer dynamical behavior within a single physical device. Chirality-controlled optoacoustic interactions may therefore offer new opportunities for compact, reconfigurable neuromorphic processors operating directly in the optical domain.

**Methods**
**1. Heterodyne setup for spontaneous Brillouin measurement**

The pump and local oscillator are derived from a narrow linewidth 1550 nm CW laser using a fiber coupler. The pump wave is then boosted in an EDFA and injected into the chiral PCF via an optical circulator. The circular polarization states are adjusted by a FPC placed before a circulator. Th polarizer, $\lambda/4$ plate and Q-plate) is optionally used to generate a CP vortex-carrying pump signal. The noise-seeded Stokes signal from the PCF is delivered by the circulator and interferes with the LO using the second 90:10 fiber coupler. Narrow-band (6 GHz) notch filters in the path of Stokes signal are used to filter out Fresnel reflections and Rayleigh scattering. The resulting beat-note is detected in the radio-frequency domain with a fast photodiode and the averaged Brillouin spectra is recorded with an electrical spectrum analyzer (ESA).

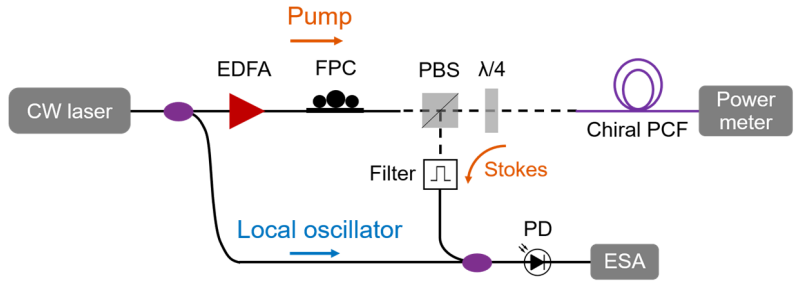

**Fig. 5** Heterodyne setup for the measurement of spontaneous Brillouin spectra.

## 2. Pump-seed setup

The pump-seed setup is shown in Figure S2. Both pump and seed were derived from a narrow linewidth 1550 nm CW laser, the seed light being frequency tuned using a single side-band modulator (SSBM). The pump signal was boosted by an EDFA and the polarization states of both pump and seed were controlled using FPCs. Vortex generating modules (circular polarizer, Q-plate and λ/2 plate) were optionally used to generate CP vortex-carrying pump signals. After propagating backwards through the chiral PCF and interacting with pump signal, the seed signal is reflected by a beam splitter (BS), filtered by a circular polarizer (polarizer and λ/4 plate) and finally delivered to NBA system for gain coefficient measurement.

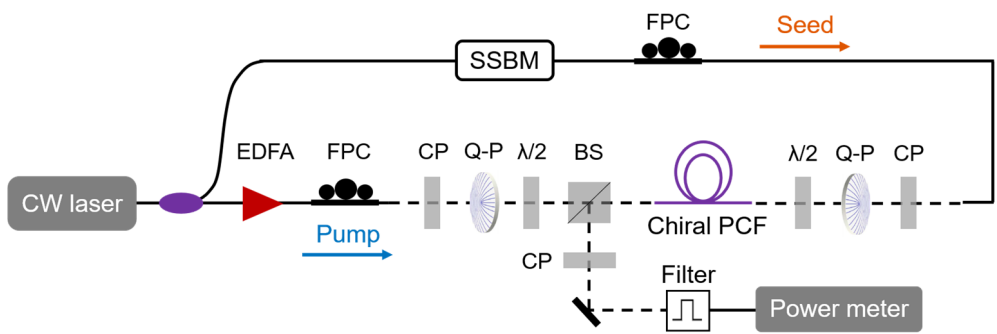

**Fig. 6.** Pump-seed experimental setup for gain coefficient measurement.

## 3. Experimental setup for the light storage of SAM-encoded single data pulse

Figure 5 shows the experimental setup of single-spin-channel light storage. A 1.5 m-long chiral PCF is used as the storage medium. A narrow-linewidth CW laser operating at 1550 nm is split into two arms: a data arm and a write–read arm. In the write-read arm, the optical signal is frequency down-shifted by the Brillouin frequency $f_B$ using a SSBM. Optical pulses in both arms are generated by IMs, which are driven by an AWG. The write-read pulses are subsequently amplified by an EDFA and then sent through a nonlinear fiber loop. This loop serves two purposes: first, it transmits only the pulsed portion while efficiently suppressing residual noise and coherent background originating from the laser and amplifiers; second, it improves the temporal pulse quality by smoothing the pulse edges. After the loop, a second EDFA further amplifies the write-read pulses to reach the required peak powers of 56 W. Two isolators are used to block the back-reflections on both arms. The FPC, PBS and λ/4 plate are used to generate two orthogonal circular polarization states on both arms. The original and retrieved data pulses are observed by a 12 GHz electrical-amplified photodiode connected to the oscilloscope. Before the photodiode a tunable narrowband filter is used to assure that only the data pulses reach the photodetector.

Two different detection schemes are used to detect the transmitted and retrieved data pulses: direct detection with a single photodiode is used for the amplitude retrieval whereas a double-homodyne detection scheme is used for the phase measurements. For the direct detection scheme the data pulses are simply detected by photodiode that connected to the oscilloscope. In the double-homodyne detection scheme, we measured the relative phase of original data and readout pulses under for different spin configuration. The 90° optical hybrid mixes the incoming signal (S) with the reference local oscillator (L) and generate four quadrature states in the complex-field space (S+L, S−L, S+jL, S−jL). Here, the signal contains initial data pulses, readout data pulses, and calibration pulses, which are used for the phase correction and constellation rotation. The local oscillator is derived from the original laser that has the same frequency as data pulse. The four signals are detected in pairs by two balanced photodiodes, enabling for differential detection of the relative phase between the S and LO, which is then mapped onto the $I(t)$ and quadrature $Q(t)$ channels. The results demonstrate the light storage is coherent, with an empirical π phase offset between the original data and the readout signals.

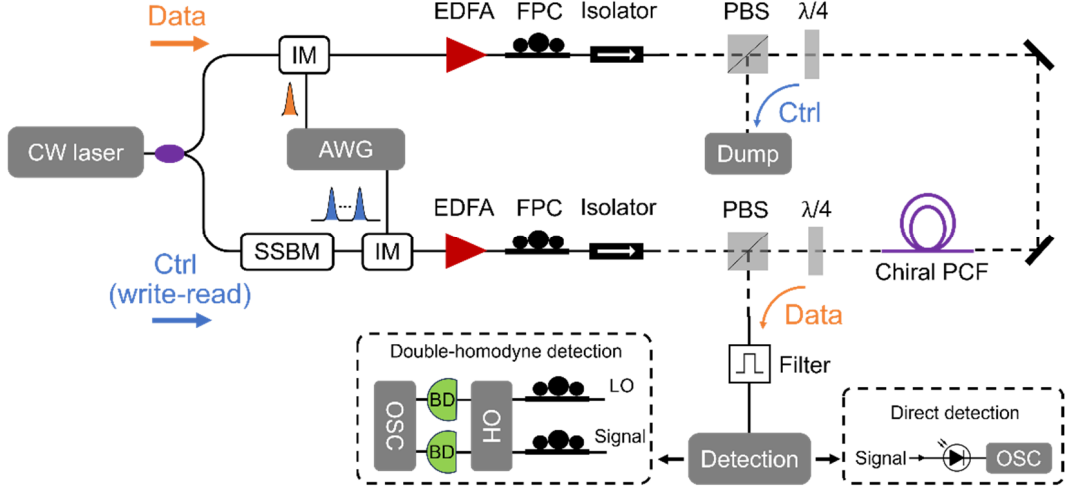

**Fig. 5** Experimental setup of SAM-encoded single data optoacoustic storage.

## 4. Experimental setup for the light storage of TSM data pulses

The experimental setup is shown in Fig. 2(c). The 1.5 m-long chiral PCF is used again as the storage medium. The output from a CW laser at 1550 nm is split into data and control paths. The control path is frequency down-shifted by $f_B$ by a SSBM. The $f_B$ is set to 10.76 GHz, corresponding to the Brillouin peak frequency of both LCP and RCP fundamental modes. An erbium-doped fiber amplifier (EDFA) is used to boost the energy of the control pulses. The data and control pulses are generated by an AWG and IM, with data and control pulse widths of 1.8 ns and repetition rate of 5 MHz. The data and write-read pulses are again boosted by high power EDFAs, with peak power of 28 W and 56 W. The TSM module is realized by splitting the data path into two channels (Data1 and Data2) through a fiber coupler, and the spin states of each channel is separately adjusted by the polarizer and $\lambda/4$ plate. A 5m- and 20 m-long SMF are used on one of the two data channels to introduce the time delay of 24.2 ns and 97 ns between them. BS2 and BS3 are used to collect the residual write–read pulses and the data pulses, respectively. The latter are subsequently filtered by a narrowband filter and detected with a high-resolution oscilloscope to characterize the writing and reading processes.

**Acknowledgements**

This research is supported by the Max Planck Research Group Scheme of the Max-Planck-Gesellschaft and the DFG project STI 792/7-1.

The authors thank Maxim Zerbib and Andreas Geilen for some preliminary tests on the photonic memory setup by using tapered nano-fibers.

**Data availability**

The data that supports the findings of this study are available from the corresponding author upon reasonable request.

**Conflict of interest**
All authors declare no conflict of interest.